\documentclass{elsart}

\usepackage{graphicx}

\usepackage{amssymb}

\begin{document}

\begin{frontmatter}



\title{Measurement of the $\Xi^-p$ 
Scattering Cross Sections at Low Energy}



\author[pnu]{J.K. Ahn\corauthref{cor1}}
\corauth[cor1]{Corresponding author: jkahn@kaon.phys.pusan.ac.kr}
\author[kobe]{S. Aoki}
\author[yonsei]{K.S.Chung}
\author[korea]{M.S.Chung}
\author[riken]{H.En'yo}
\author[osakaec]{T.Fukuda}
\author[kyoto]{H. Funahashi}
\author[riken]{Y. Goto}
\author[ins]{A. Higashi}
\author[kek]{M. Ieiri} 
\author[kek]{T. Iijima}
\author[kyoto]{M. Iinuma\thanksref{hiroshima}}
\author[kyoto]{K. Imai} 
\author[icrr]{Y. Itow}
\author[yonsei]{J.M. Lee\thanksref{kriss}}
\author[kyoto]{S. Makino} 
\author[kyoto]{A. Masaike} 
\author[kyoto]{Y. Matsuda}
\author[ins]{Y. Matsuyama}
\author[kyoto]{S. Mihara}
\author[ins]{C. Nagoshi}
\author[nifs]{I. Nomura}
\author[korea]{I.S. Park}
\author[kyoto]{N. Saito}
\author[kek]{M. Sekimoto}
\author[saska]{Y.M. Shin}
\author[korea]{K.S. Sim}
\author[riken]{R. Susukita}
\author[kyotoed]{R. Takashima} 
\author[kyotosg]{F. Takeutchi}
\author[ins]{P. Tlust$\rm{\acute{y}}$\thanksref{czech}}
\author[saska]{S. Weibe}
\author[kyoto]{S. Yokkaichi}
\author[kyoto]{K. Yoshida}
\author[riken]{M. Yoshida}
\author[osakac]{T. Yoshida} 
\author[icep]{S. Yamashita}

\address[pnu]{Department of Physics, Pusan National
University, Pusan, 609-735, Korea}
\address[kobe]{Faculty of Human Development, Kobe University, Kobe 657-8501, Japan}
\address[yonsei]{Department of Physics, Yonsei University, Seoul 120-749, Korea}
\address[korea]{Department of Physics, Korea University, Seoul 136-701, Korea}
\address[riken]{RIKEN, Institute for Physical and Chemical Research, 
Wako, 351-0198, Japan}
\address[osakaec]{Research Center for Physics and Mathematics, Osaka Electro-Communication University, Osaka, 572-8530, Japan}
\address[kyoto]{Department of Physics, Kyoto University, Kyoto 606-8502, Japan}
\address[ins]{Institute for Nuclear Study, University of Tokyo, Tokyo 188-8501, Japan}
\address[kek]{IPNS, KEK, High Energy Accelerator
Research Organization, Oho, 1-1, Tsukuba 305-0801, Japan}
\address[icrr]{Institute for Cosmic Ray Research, 
University of Tokyo, Tokyo 188-8502, Japan} 
\address[nifs]{National Institute for Fusion Science, Nagoya 464, Japan}
\address[saska]{University of Saskatchewan, Saskatoon, Canada}
\address[kyotoed]{Kyoto University of Education, Kyoto 612, Japan}
\address[kyotosg]{Kyoto Sangyo University, Kyoto 606, Japan}
\address[osakac]{Department of Physics, Osaka City University, Osaka 558-8585, Japan}
\address[icep]{International Center 
for Elementary Particle Physics, University of Tokyo, Tokyo 113, Japan}

\thanks[hiroshima]{Present address : Department of Physics, Hiroshima University, 
Hiroshima 739-8526, Japan}
\thanks[kriss]{Present address : Korea Research Institute for Standards
and Science, Dajeon 305-600, Korea} 
\thanks[czech]{Present address : Nuclear Physics Institute of the Academy
of Science, 250 68 $\rm{\check{R}}$e$\rm{\check{v}}$, $\rm{\check{C}}$zech Republic}

\begin{abstract}

In this paper we report cross-section measurements for $\Xi^-p$
elastic and inelastic scatterings at low energy using a scintillating
fiber active target. Upper limit on the total cross-section for the elastic
scattering was found to be 24 mb at 90\% confidence level, and the
total cross section for the $\Xi^-p\rightarrow\Lambda\Lambda$ reaction
was found to be $4.3^{+6.3}_{-2.7}$ mb. 
We compare the results with currently competing
theoretical estimates. 

\end{abstract}


\end{frontmatter}


Hyperon-nucleon interaction, particularly in $S=-2$ system, has
attracted great attention since it may shed light on our
understanding of the nature of SU(3) symmetry breaking in a
baryon-baryon interaction.
Its importance is more highlighted to account for a possible
six-quark $H$-dibaryon, which has been as yet unobserved
experimentally. In addition, the $\Xi$N-$\Lambda\Lambda$ 
interaction accounts for the existence of a
double hypernucleus, which is a gateway to a strange hadronic matter.
For elementary $\Xi^-p$ scattering, there is a scanty supply of
bubble-chamber experimental data at higher energies above 
1 GeV/{\it c} \cite{charlton,dalm,muller,haupt}.

Since no elementary
cross sections below 1 GeV/{\it c} are available to date, 
our attempt to measure
the $\Xi^-p$ elastic scattering and conversion processes is
very important in testing competing theoretical-model predictions 
\cite{rgmf,fss,rgmh,nakamoto,su6}.
In particular, a measurement of the $\Xi^-p\rightarrow\Lambda\Lambda$ 
conversion 
cross section is crucial in assessing the stability of $\Xi^-$
quasi-particle states in nuclei.


Our experiment was carried out using a separated 
1.66 GeV/{\it c} $K^-$ beam at the KEK proton synchrotron. 
The momenta of outgoing particles were measured with a $K^+$
spectrometer of approximately 5 m length.
The momentum resolution ($\Delta p/p$) was 0.5\% (rms) 
at 1.2 GeV/{\it c}, and the background ratio of $\pi^+$ to $K^+$ was
estimated to be less than 0.2 \%. 
The scintillating fiber (SCIFI) detector 
consisting of $0.5\times 0.5$mm$^2$ plastic scintillating fibers 
acts as a primary interaction
target and visual track detector for identifying
$(K^-,K^+\Xi^-)$ reactions, and secondary interactions of $\Xi^-$
in the effective volume of $8\times 8\times 10$ cm$^3$.
The $\Xi^-$ particles are produced by $(K^-,K^+)$ reactions on
free protons or bound protons in carbon nuclei. 
The SCIFI detector was read using two sets of the image-intensifier tubes
(IIT), which were arranged orthogonally. 
The track residual, defined as the distribution of the distance 
between the fitted straight track and each cluster, weighted by the
brightness of each cluster, was about 290 $\mu$m.
The detailed description of the experimental methods appears in
elsewhere \cite{ahn1,phd,mschung}.

\begin{figure}[hbt]
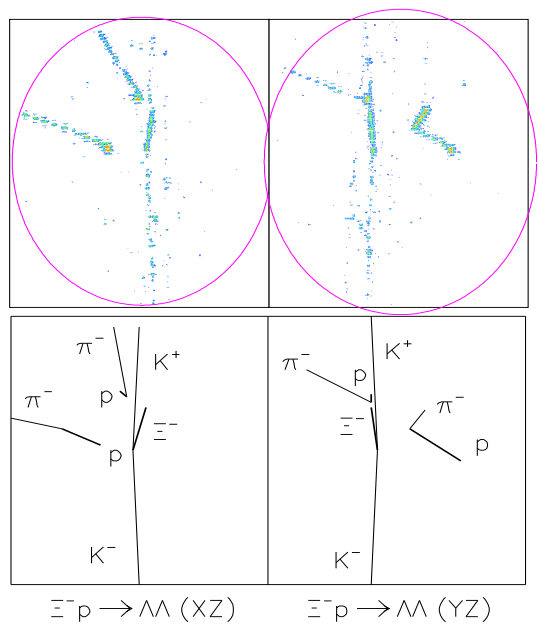

\centering
\begin{minipage}[t]{3.1in}
\includegraphics[height=1.6in,width=7.cm]{xpll3.epsi}
\end{minipage}
\begin{minipage}[t]{3.1in}
\includegraphics[height=1.6in,width=7.cm]{xp4conf.epsi}
\end{minipage}
\caption{Event sample for 
$\Xi^-p\rightarrow\Lambda\Lambda$ 
scattering process with sketch images}
\label{fig:xpimage}
\end{figure}

The event configuration of $\Xi^-p$ elastic scattering is characterized as 
two particle tracks from the $\Xi^-$ interaction vertex and one of the two
involving a kink. 
On the other hand, the $\Xi^-p\rightarrow\Lambda\Lambda$ reaction 
candidates were identified by observing two V topology tracks and one stopping
track from the $(K^-,K^+)$ primary vertex, as shown in Fig. \ref{fig:xpimage}.

Out of over six thousand $(K^-,K^+)\Xi^-$ events 
a total of two candidate events were selected
as $\Xi^-p$ elastic scattering.  
The $\Xi^-p$ candidates were selected by requiring
that the track length of a recoil proton be longer than 3 mm and that of a
scattered $\Xi^-$ be also longer than 3 mm. None of them has a visible
$\Lambda$ decay following the decay of a scattered $\Xi^-$:
$\Xi^-\rightarrow\Lambda\pi^-$ and $\Lambda\rightarrow p\pi^-$.  
After imposing constraints on two-body $\Xi^-p\rightarrow\Xi^-p$ scattering,
one of the two finally survived. 

We also observed three candidate events involving two visible
$\Lambda$ decays from a stopping prong at the $(K^-,K^+)$ vertex.
The three events contain two $\Lambda$ particles 
carrying unbalanced momenta in terms of two-body kinematics;
the residual nucleus carries the unbalanced
missing momentum. We therefore interpreted these events as due to the reaction
$^{12}$C$(\Xi^-,\Lambda\Lambda)$X. 

%
\indent
A great deal of attention was paid to estimate detection efficiencies.
The detection efficiencies were calculated by a Monte-Carlo simulation based on
{\sc GEANT}. The incident $\Xi^-$ particles were generated to reproduce
the measured $\Xi^-$ momentum spectrum, and scattered off free
protons. Energy loss in the SCIFI target medium was taken into
account. The simulation involves 
reproducing the response of the SCIFI detector through hit digitization
according to the individual fiber and fiber sheet configurations. 
The simulated $\Xi^-p$ scattering events were scanned manually by eye
in the same way as the analysis of real data. 
The scanning efficiency was studied as a function of track length and
opening angle.


The acceptance was estimated by using a large number of
the simulated events in terms of the $\Xi^-$ momentum and c.m. angle of
scattering.
Proper cutoffs were imposed on the track length and opening angle
according to the scanning results of the simulated events. 
The track length for all the charged particles was required to
be longer than 3 mm and the $\Lambda$ flight length longer than 5 mm.
The opening angle between the scattered $\Xi^-$ and the
recoil proton should be larger than 
$\cos\theta_{\Xi p}=$0.96. 

\indent
We calculate upper limit on cross section for $\Xi^-p$ elastic 
scattering based on the observation of 
one $\Xi^-p$ elastic scattering event with an invisible $\Lambda$ decay.
If only incident and scattered
$\Xi^-$ particles are taken into account regarding $\Xi^-p$
elastic scattering, the acceptance is flat over
the range of c.m. scattering angles. However, as we discussed above, 
with the requirement of the visible $\Lambda$ decay
the acceptance is not flat in very forward
and backward directions. We therefore obtain the differential cross section
for the elastic scattering channel in the range of the $\Xi^-$ scattering angle
$-0.35<\cos\theta^{cm}_{\Xi^-p}<0.65$. 

\indent
The differential cross sections for $\Xi^-p$ scattering processes
can be written as
$2\pi(\frac{d\sigma}{d\Omega})\Delta\cos\theta =\frac{N_{\Xi^-p}/\eta_{\Xi^-p}}
{\rho \cdot L^{\rm tot}_{\Xi^-}},$
where $N_{\Xi^-p}$ denotes the number of $\Xi^-p$ scattering events.
The target density of free protons, $\rho=4.88\times 10^{22}$ cm$^{-3}$, 
can be rewritten as $\rho_0N_A/W$, where
$\rho_0$ is the density of the target material (CH)$_n$ in g/cm$^3$, 
$N_A$ is the Avogadro's
number, and the atomic weight, $W$, is given by 13 g/mol
The detection efficiency, $\eta_{\Xi^-p}$ was obtained by a 
Monte-Carlo simulation, as described in the previous paragraph. 
The total track length of the incident $\Xi^-$
particles is represented by the factor
$L^{\rm tot}_{\Xi^-}= \sum_{i=1}^{N^{\rm tot}_{\Xi^-}} ( L^{\Xi^-}_i -L_0) =
\frac{N^{\prime}_{\Xi^-}}{\eta_{\Xi^-}} \cdot 
(\sum L^{\Xi^-}_i/N^{\prime}_{\Xi^-} - L_0)$,
where $N^{\rm tot}_{\Xi^-}$ denotes the total number of $\Xi^-$
particles and $N^{\prime}_{\Xi^-}=3834$ indicates the analyzed number of 
$\Xi^-$ particles. $L^{\Xi^-}_i$ stands for the track length of each
$\Xi^-$. The effective target length of
the analyzed event set was estimated to be 2.39 cm after correction
due to the SCIFI target geometry.
The factor $\eta_{\Xi^-}$ denotes the ratio of
the number of analyzed $\Xi^-$ events to the total number of
$\Xi^-$ events, which was found to be 60.1\%. 
The minimum cutoff length of $\Xi^-$ track, $L_0$, is 0.3 cm. 
The total track length of the incident $\Xi^-$ particles, 
$L^{\rm tot}_{\Xi^-}$, was then found to be
$1.34\times 10^4$ cm. 

\begin{figure}[h]
\begin{center}
\includegraphics[height=8cm,width=8cm]{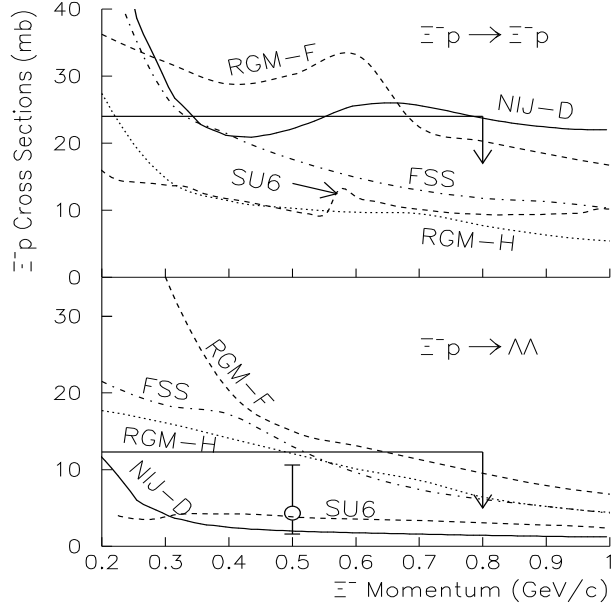}
\caption{Upper limits on the $\Xi p$ cross sections at 90\% confidence 
level, indicated by arrows,
are compared with theoretical estimates of RGM-F, RGM-H, FSS, 
Nijmegen-D, and SU$_6$ quark model. In the bottom panel a data point represents
the result obtained from $\Xi^{-12}$C$\rightarrow\Lambda\Lambda$
reaction assuming the effective proton number of 3.5. Poisson
statistical error is quoted only.} 
\label{fig:xpres}
\end{center}
\end{figure}

Assuming Poisson statistics, we set 
upper limits on the cross sections at 90\% confidence
level (2.44 for 0 event and 3.86 for 1 event with 50\% background).
The differential cross section for the $\Xi^-p$ elastic
scattering can be obtained in two ways: one is based on 
the observation of an event involving no $\Lambda$ decay, the other
is due to no observation of such an event that involves a
visible $\Lambda$ decay following the scattered $\Xi^-$ decay.
The detection efficiency factors $\eta_{\Xi^-p}$ are
20.4\% and 43.9\%, respectively. 
For the calculation based on an observed event, we take 0.5 
background events in the signal, simply because of taking
the similar probability of quasifree elastic scattering from carbon nuclei 
into account. 
Upper limit on the differential cross sections 
for the elastic scattering channel in the range of
$-0.35<\cos\theta^{cm}_{\Xi^-p}<0.65$ were then found to
be 1.9 mb/sr (with $\Lambda$) and 2.1 mb/sr (without $\Lambda$)
at 90\% confidence level, respectively. 
The null result for the $\Xi^-p$ elastic scattering with
$\Lambda$ decay provides more strict upper limit on the 
elastic scattering cross section.
For simplicity, assuming
an isotropic angular distribution, upper limit on the total cross section for
$\Xi^-p\rightarrow\Xi^-p$ scattering process is 24 mb 
at 90\% confidence level, as shown in Fig.\ref{fig:xpres}. 

Regarding the $\Xi^-p
\rightarrow\Lambda\Lambda$ reaction, we calculated the total cross-section with
a null result. The detection efficiency was found to be 74\% 
over the whole region of c.m. angles, and the branching ratio of
two $\Lambda\rightarrow p\pi^-$ decays (41\%) was taken into account.
Upper limit on the total cross-section for
$\Xi^-p\rightarrow\Lambda\Lambda$ process was obtained to be 12 mb at
90\% confidence level. We also calculated the total cross section
based on three observed events due to the reaction
$^{12}$C$(\Xi^-,\Lambda\Lambda)$X. We simply assumed a quasifree 
scattering process. The effective proton number for the 
$^{12}$C$(\Xi^-,\Lambda\Lambda)$X reaction reflects the initial
state interaction of $\Xi^-$ through the $\Xi^-p\rightarrow\Xi^0n$ 
process and the sticking probability that either of two $\Lambda$
particles or both stays inside a nucleus, thereby decaying weakly.
In eikonal approximation, the effective nucleon number can be written as
\begin{eqnarray} 
N_{(\Xi^-,\Lambda\Lambda)}=(1-\overline{S}_{\Lambda})^2
\int_0^{\infty} 2\pi b{\rm d}b \int_{-\infty}^{\infty} {\rm d}z
\rho(\sqrt{b^2+z^2}) \nonumber \\ 
\times\mbox{exp}\biggl[-\bar{\sigma}_{\scriptsize
\Xi^-p\rightarrow\Xi^0n} \int^z_{-\infty}
{\rm d}z'\rho(\sqrt{b^2+z'^2})\biggr], \nonumber
\end{eqnarray}
where atomic mass $A=\int d\vec{r}\rho(\vec{r})$, where the nuclear radius 
$r$ is given with the impact parameter $b$ and the $z$ coordinate.  
The nuclear density is of
three parameter Fermi-distribution of 
$\rho(r)=(1-wr^2/r_c^2)/(1+\mbox{exp}(r-r_c/a))$, where $w=0.149$, $a=0.5224$,
and $r_c=2.355$ fm. 
The $\Lambda$ sticking probability was assumed to be 
$\overline{S}_{\Lambda}=0.15$\cite{sticking}. 
The effective proton number 
$Z_{\rm eff}=(Z/A)\times N_{\rm eff}$ is then found 
to range from 3.0 to 4.1 with 
$\sigma_{\Xi^-p\rightarrow\Xi^0n}= 18$ mb (predicted by RGM models 
\cite{rgmf,fss,rgmh})
and 2 mb (by Nijmegen-D model \cite{nakamoto}), respectively.
The SU$_6$ quark model predicts $\sigma_{\Xi^-p\rightarrow\Xi^0n}$ to
be about 15 mb at $p_{\Xi}=0.5$GeV$/c$ \cite{su6}. 
Our previous experimental result ($\approx 14$ mb)
for inelastic scattering channels
involving both $\Xi^-p\rightarrow\Lambda\Lambda$ and
$\Xi^-p\rightarrow\Xi^0n$ processes suggests that
$\sigma_{\Xi^-p\rightarrow\Xi^0n}$ be of the order of 10 mb\cite{ahn4}.
For $\sigma_{\Xi^-p\rightarrow\Xi^0n}=10$ mb,
the effective proton number was obtained to be 3.5. This value 
in turn gives the total cross-section for
$\Xi^-p\rightarrow\Lambda\Lambda$ of $4.3^{+6.3}_{-2.7}$ mb
(statistical error only).


\indent
We discuss possible systematic uncertainties in estimating the
cross sections for $\Xi^-p$ elastic scattering and conversion processes.
The $\Xi^-p$ cross sections measured in different ways are consistent
with each other, which permit us to draw reliable
conclusions
on the low-energy $\Xi^-p$ and $\Lambda\Lambda$ interactions.
Special attention was paid to quantify the systematic uncertainties.
The possible uncertainties are discussed as follows:
(1)The total cross section measurement
for the $\Xi^-p$ elastic scattering was based on the assumption
of an isotropic angular distribution. If one takes the distribution
$d\sigma/d\cos\theta^{cm}_{\Xi^-p}$ $\sim 1+\cos^2\theta^{cm}_{\Xi^-p}$, 
then the total cross section would be 20\% larger than the estimate
based on an isotropic angular distribution.
(2) Uncertainty of the detection efficiency was
 investigated by varying cutoffs on the minimum track length.
It was found to be 5\% change in the cross section results. 
(3) The SCIFI target is composed of almost equal amount of hydrogen
and carbon, and also includes a tiny amount of other atomic components
due to cladding and glue materials. 
This contribution was estimated to
be of the order of 1\% in the cross section results. 
It should be also noted that all the criteria for 
event selection were determined by scanning and analyzing a number of 
simulated events.
All the systematic uncertainties are much smaller than limited 
significance of our statistics. 
  
The total cross section for the $\Xi^-p\rightarrow\Lambda\Lambda$ 
conversion process sets tight limit on theoretical estimates
as shown in Fig. \ref{fig:xpres}. The upper limit on the cross section, 
12 mb at 90\% confidence level, rules out the RGM-F
model prediction \cite{rgmf}. The measured cross section based on quasifree
assumption is $4.3^{+6.3}_{-2.7}$ mb, which supports rather the prediction
of the model D(r$_c=0.5$ fm) \cite{nakamoto} 
of Nijmegen group and the SU$_6$ quark model \cite{su6}.
Note that the quasifree 
scattering cross section depends on the estimate of the effective
number of protons involved in the $^{12}$C$(\Xi^-,\Lambda\Lambda)X$ 
reaction. The above estimate was based on the effective proton
number of 3.5, as discussed earlier. 
Due to this uncertainty of the effective number 
of protons, the models FSS \cite{fss} and
RGM-H \cite{rgmh} cannot be ruled out.

\begin{figure}[thb]
\begin{center}
\includegraphics[height=6cm,width=8cm]{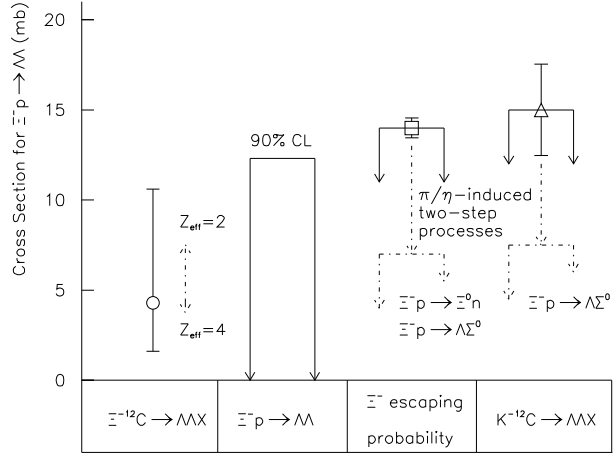}
\caption{Cross sections for the $\Xi^-p\rightarrow\Lambda\Lambda$
reaction with respect to different reaction channels: (from left)
$\Xi^{-12}$C$\rightarrow\Lambda\Lambda$X with three observed events,
$\Xi^-p\rightarrow\Lambda\Lambda$ with a null result, 
$^{12}$C$(K^-,K^+)\Xi^-$X, and $^{12}$C$(K^-,K^+\Lambda\Lambda)$X with
intra-nuclear cascade model prediction for intermediate $\Xi^-$ and
meson-induced processes.}
\label{fig:xrosll}
\end{center}
\end{figure}

We discuss consistency of $\Xi^-p\rightarrow\Lambda\Lambda$
cross sections measured using four different reactions;
$\Xi^-p\rightarrow\Lambda\Lambda,
\Xi^{-12}{\rm C}\rightarrow\Lambda\Lambda X,
K^{-12}{\rm C}\rightarrow K^+\Lambda\Lambda X,$
and $K^{-12}{\rm C}\rightarrow K^+X$ (no $\Xi^-$).
Among them the direct measurement of elementary
$\Xi^-p\rightarrow\Lambda\Lambda$
process should be emphasized since it is free from theoretical assumptions. 
Despite the measurement
yielded no observed event, it sets the stringent 
upper limit on the cross section to be 12 mb at 90\%
confidence level. Secondly, quasifree
$\Xi^{-12}{\rm C}\rightarrow\Lambda\Lambda$X reaction 
events involving two momentum-unbalanced $\Lambda$ particles
permit us to obtain the cross section of
$4.3^{+6.3}_{-2.7}$ mb. The errors quoted indicate 
a 90\% confidence level interval for Poisson signals.
This result basically contains an assumption of the effective
proton number being 3.5 in the eikonal approximation. The uncertainty
in estimating the effective proton number is indicated as
dashed-dotted arrow lines in the same column of Fig. 
\ref{fig:xrosll} (3.8 mb for $Z_{\rm eff}=4$ and 7.5 mb for 
for $Z_{\rm eff}=2$). 

We also estimated the total cross section from
the reactions, $K^{-12}{\rm C}\rightarrow K^+\Lambda\Lambda$X
and $K^{-12}{\rm C}\rightarrow K^+$X without $\Xi^-$ production.
These processes however compete with intermediate meson-induced
two-step processes \cite{ahn4}. Both cases could also involve rather
sizable contribution from $\Xi^-p\rightarrow\Lambda\Sigma^0$
due to Fermi motion of nucleons in $^{12}$C above 0.57 GeV/$c$.   
The ratio of the observed number of the $\Xi^-$
particles to that of the $^{12}$C$(K^-,K^+)$ reaction events would 
provide upper limits on the inelastic $\Xi^-N$ cross section. 
The ratio was found to be 
$80.1 \pm 2.4^{+0.6}_{-1.3}$\% \cite{mschung}. The total cross section for
$\Xi^-p$ inelastic scattering was then deduced from the $\Xi^-$ 
absorption probability for $^{12}$C,
which is approximately 14 mb assuming an isotropic angular distribution.
A recent measurement of quasifree $p(K^-,K^+)\Xi^-$ reaction in emulsion plate
yields $12.7^{+3.5}_{-3.1}$ mb in the momentum range 0.4--0.6 GeV/{\it c}
\cite{yasuda}, which is well consistent with our result.
On the other hand, the analysis of 
the $\Lambda\Lambda$ invariant mass spectrum 
suggests that the $\Xi^-p\rightarrow\Lambda\Lambda$ cross section 
is of the order of several mb at $p_{\Xi^-}\sim 0.55$ GeV/{\it c} \cite{ahn3}.

In nuclear matter the $\Xi^-$ decay width $\Gamma_{\Xi^-}\simeq
(v\sigma)_{\Xi^-p\rightarrow\Lambda\Lambda}
\cdot(\rho^0/2)$ \cite{dover} arrives at $\sim 3$ MeV, based on the result
of $4.3^{+6.3}_{-2.7}$ mb. 
The term $\rho^0/2 \sim 0.08$ fm$^{-3}$ denotes 
proton density and $v$ is a $\Xi^-p$ relative velocity. 

In summary the KEK-PS measurement provides the first experimental data
on low-energy $\Xi^-p$ elastic and conversion cross sections in the range
from 0.2 GeV$/c$ to 0.8 GeV$/c$, which
are less than 24 mb at 90\% confidence level and of the order of
several mb, respectively. The four independent and consistent 
measurements of $\Xi^-p\rightarrow\Lambda\Lambda$
cross section rule out a theoretical estimate based on the RGM-F model
calculation, and favor the estimates predicted by Nijmegen 
model-D(r$_c=0.5$ fm) and SU$_6$ quark model. In nuclear matter, 
the width for a $\Xi^-$ single-particle state is estimated to be 
$\Gamma_{\Xi^-}\approx 3$ MeV.

We would like to thank the staffs of the KEK proton synchrotron for
their continuous support. We also thank Prof. Y. Fujiwara, 
Dr. C. Nakamoto and Dr. Y. Nara for valuable discussions
and for providing us with theoretical calculation results. 
This work was supported in part by a Grant-in-Aid 
for Scientific Research (08239103), the Ministry 
of Science, Culture and Education, Japan. The work of J.K. Ahn was supported 
by Korea Research Foundation grant (KRF-2003-015-C00130).


\end{document}